\documentclass[final]{mem}
\usepackage{amsmath, amssymb}
\usepackage[T1]{fontenc}
\usepackage[utf8]{inputenc}
\usepackage{natbib}
\usepackage[varg]{txfonts}
\usepackage{graphicx}
\usepackage[a4paper,breaklinks,dvipdfm]{hyperref}
\usepackage[english]{babel}
\begin{document}
\newcommand*{\mstar}{M_\star}
\newcommand*{\msun}{M_\odot}
\newcommand*{\Mh}{M_h}
\newcommand*{\Mtot}{M_\mathrm{tot}}
\newcommand*{\Mhot}{M_\mathrm{hot}}
\newcommand*{\Mcold}{M_\mathrm{cold}}
\newcommand*{\Minj}{M_\mathrm{inj}}

\newcommand*{\Tbol}{\langle T \rangle_\mathrm{bol}}
\newcommand*{\Tx}{\langle T \rangle_\mathrm{X}}
\newcommand*{\Tmass}{\langle T \rangle_\mathrm{mass}}

\newcommand*{\Lbol}{L_\mathrm{bol}}
\newcommand*{\Lx}{L_\mathrm{X}}
\newcommand*{\Lsn}{L_\mathrm{SN}}
\newcommand*{\Lkin}{L_\mathrm{kin}}
\newcommand*{\Lstr}{L_\mathrm{str}}
\newcommand*{\Lsigma}{L_{\sigma}}
\newcommand*{\vphi}{\overline{v}_{\varphi}}
\newcommand*{\vphisqmean}{\overline{v_\varphi^2}}
\newcommand*{\sigmaphi}{\sigma_\varphi}


\title{Disk dynamics and the X-ray emission of S0 and flat early-type galaxies}
\subtitle{}
\author{
A. \,Negri\inst{1},
S. \,Pellegrini\inst{1}, and L. \,Ciotti\inst{1}}
\offprints{A. Negri}
\institute{Dipartimento di Fisica e Astronomia, Università di Bologna, viale Berti Pichat 6/2, 40127 Bologna, Italy;
\email{andrea.negri@unibo.it}
}

\authorrunning{Negri et al.}
\titlerunning{Disk dynamics and the X-ray emission of S0 and flat early-type galaxies}

\abstract{With 2D hydrodynamical simulations, we study the evolution of the hot gas flows in early-type
galaxies, focussing on the effects of galaxy rotation on the thermal and dynamical status of the ISM. The
galaxy is modelled as a two-component axisymmetric system (stars and dark matter), with a variable amount of azimuthal
velocity dispersion and rotational support; the presence of a counter rotating stellar disk is also considered.
It is found that the ISM of the rotationally supported (isotropic) model is more prone to thermal instabilities than the fully velocity dispersion counterpart, while its ISM temperature and X-ray luminosity are lower.
The model with counter rotation shows an intermediate behaviour.
}
\maketitle{}

\section{Introduction}
\begin{figure*}[t!]
\resizebox{0.331\linewidth}{!}{\includegraphics[clip]{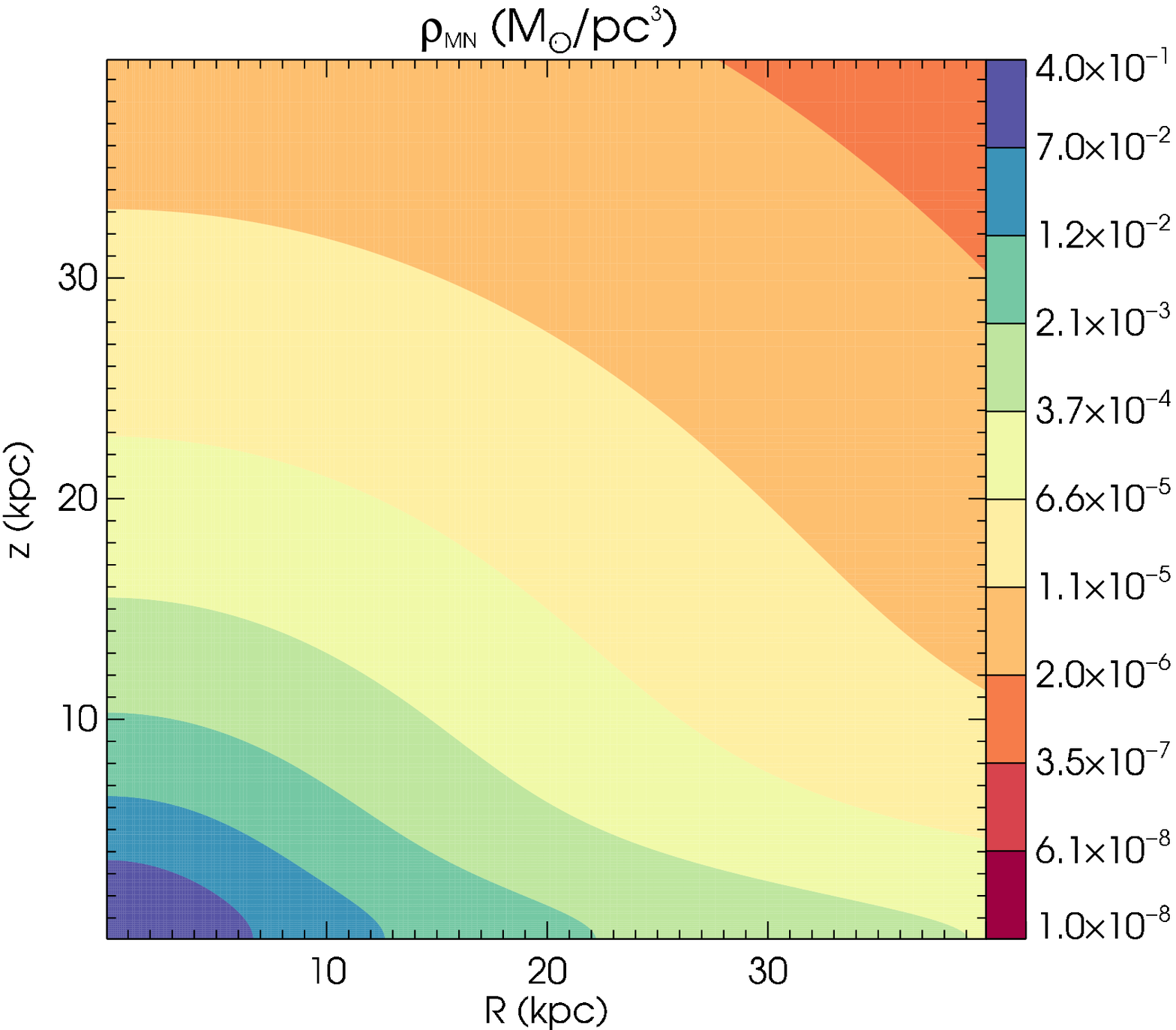}}
\resizebox{0.325\linewidth}{!}{\includegraphics[clip]{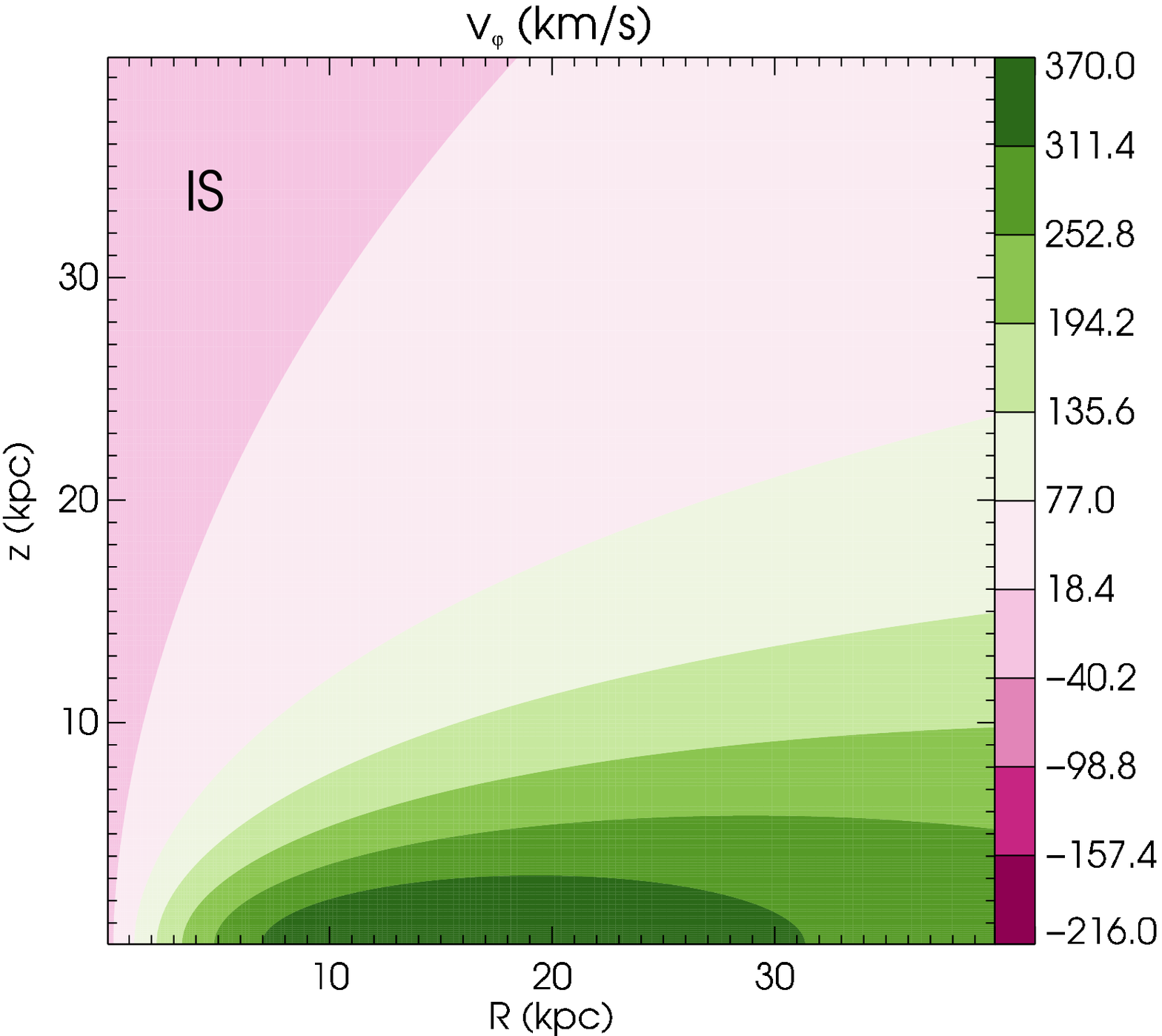}}
\resizebox{0.325\linewidth}{!}{\includegraphics[clip]{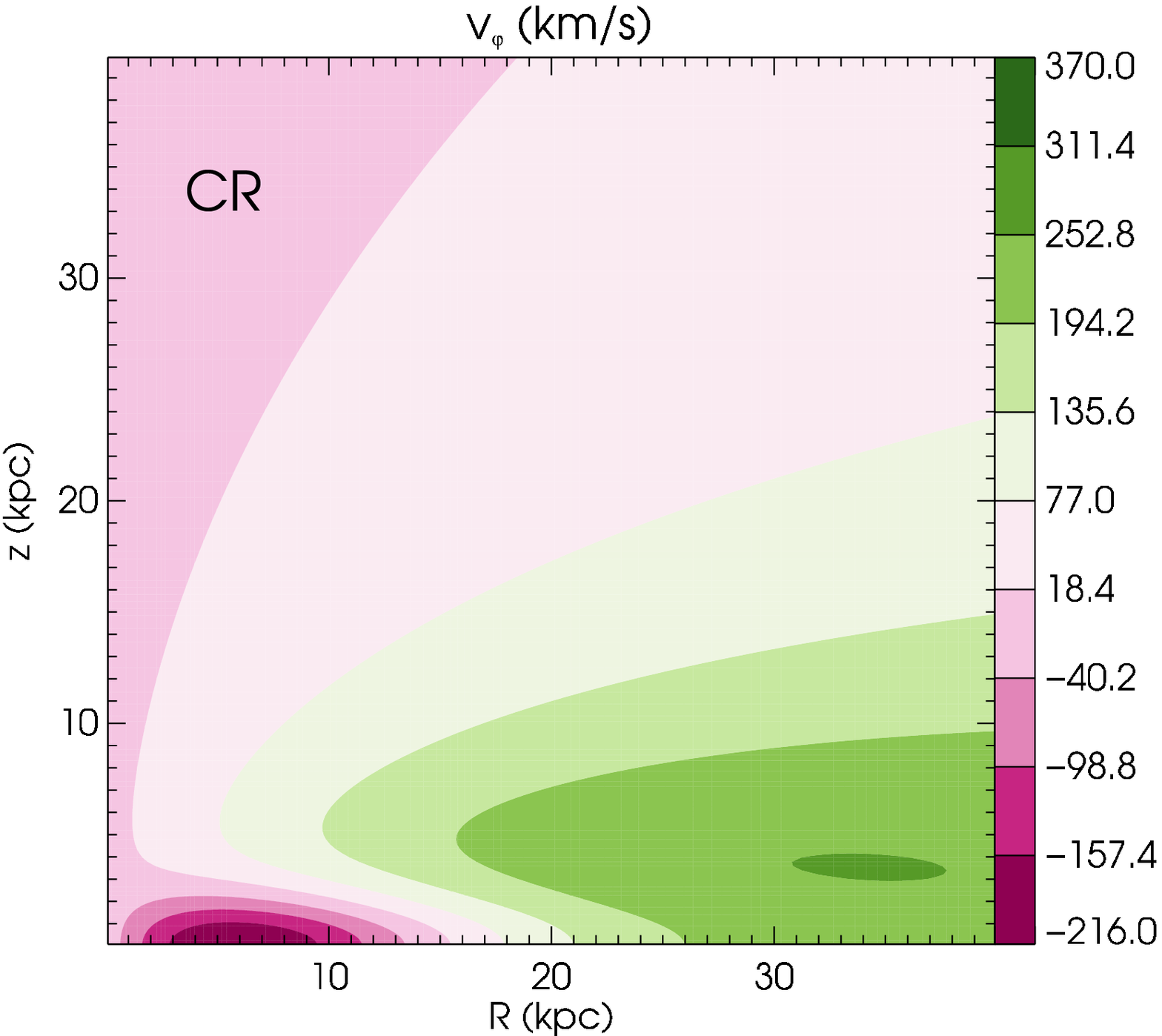}}
\caption{\footnotesize Meridional sections of the stellar density (left) and $\vphi$ for the IS (middle),
and CR (right) cases.}
\label{fig1}
\end{figure*}

The hot interstellar medium (ISM) of early-type galaxies
is observed to be sensitive to the galactic shape, and possibly also
to the stellar kinematics \citep{P12}.
In order to clarify the situation from the theoretical point of view, we have undertaken a numerical investigation of the hot gas evolution in flat galaxies, focussing on the effect of different rotational fields.  
We adopted an axisymmetric Miyamoto-Nagai stellar distribution $\rho_\mathrm{MN}$, with $a=b=3$ kpc, and $\mstar = 2.8 \times 10^{11} \msun$, immersed in a dark matter halo described by a Plummer model.
The ordered ($\vphi$) and random ($\sigma^2$ and $\sigmaphi^2$) stellar velocities are obtained from the Jeans equations and the Satoh decomposition with parameter $k$ \citep{binney2008}.
%
Three cases are considered: the isotropic rotator (IS, $k=1$), the
fully velocity dispersion supported case (VD, $k=0$), and the counter rotating disk
case (CR), modelled with space dependent Satoh parameter
\begin{equation}
 -1 \leq k= 1-2 \rho_{\mathrm{MN}} (R,z)/\rho_{\mathrm{MN}} (0,0) \leq 1. \label{conter_rot}
\end{equation}
In Eq.~(\ref{conter_rot}) we adopt $a=18$ kpc, $b=4$ kpc, so that $\vphi$ counter-rotates inside a very thin Miyamoto-Nagai disk, while outside the velocity field approaches that of the IS model (Fig.~\ref{fig1}).
The simulations include mass sources for the hot ISM as predicted by stellar evolution,
heating from the thermalization of stellar motions and type Ia supernova explosions.
We recall that while the thermalization of $\sigma$ is independent of the local ISM velocity,
the degree of thermalization of $\vphi$ depends on the relative motion between the fresh gas injected and the pre-existing ISM, and so it cannot be estimated a priori.
Interesting effects should take place if a rotating cooling flow collapses onto a counter rotating disk, with the associated mass and energy sources: a reduction of the net angular momentum of the rotating inflow is expected, with an enhanced tendency to collapse.
However, a competing effect is also present, i.e. the extra-thermalization of the counter rotating mass sources.


\section{Results}
\begin{figure}[]
\resizebox{\linewidth}{!}{\includegraphics[clip]{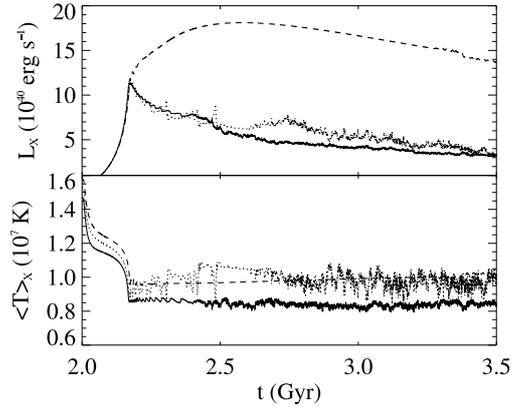}}
\caption{\footnotesize Time evolution of the ISM X-ray luminosity and X-ray luminosity weighted temperature for the IS (solid), VD (dashed) and CR (dotted) models.}
\label{fig2}
\end{figure}

The ISM evolution is followed with ZEUS-MP2 in cylindrical coordinates $(R,z)$ with 200x400 grid points, and a spatial resolution of 200 pc.
After the density and temperature reach critical values, a central ``cooling flow'' develops (at $\simeq 2.2$ Gyr, Fig.~\ref{fig2}), and the flow decouples in a cold ($T<10^4~K$) and dense disk (in the IS and CR), $\simeq 2$~kpc wide (smaller in CR), embedded in a rarefied hot ($T>10^6$~K) outflowing atmosphere.
Minor cooling episodes, i.e., gas accretion events on the central cold disk, continuously take place.
%
%
The IS model is the coldest one, with the lowest bolometric and X-ray luminosities and hot gas content, due to the significant centrifugal support and the lack of thermalization of the ordered motions. 
The VD model is the hottest and X-ray brightest model, with the lowest content of cold gas ($\simeq 1/2$ of IS). In fact, all stellar motions are thermalized, and a single, major cooling episode leaves a dense core instead of a disk, due to the lack of angular momentum of the mass sources.
The CR case is marginally hotter (and more brighter, both in the bolometric and X-ray bands) than the IS, due to the additional thermalization provided by the counter rotating region, and its surrounding velocity dispersion supported zone.
However, both models have a similar content of hot and cold gas.

Finally, we found that the efficiency of thermalization of the ordered stellar motions is quite small, with important implications for studies of the relation between gas temperature and internal kinematics (Posacki at al., this volume).

\bibliographystyle{aa}

\end{document}